\documentclass{article}

\usepackage{smc}
\usepackage{times}
\usepackage{ifpdf}
\usepackage[english]{babel}
\usepackage{csquotes}
\usepackage[natbib=true,style=numeric,sorting=none]{biblatex}
\def\papertitle{ESTIMATING THE REPERTOIRE SIZE IN BIRDS USING UNSUPERVISED CLUSTERING TECHNIQUES}
\def\firstauthor{Joachim Poutaraud}

\newif\ifpdf
\ifx\pdfoutput\relax
\else
   \ifcase\pdfoutput
      \pdffalse
   \else
      \pdftrue
\fi

\ifpdf 
  \usepackage[pdftex,
    pdftitle={\papertitle},
    pdfauthor={\firstauthor},
    bookmarksnumbered, 
    pdfstartview=XYZ 
   ]{hyperref}

  \usepackage[pdftex]{graphicx}
  \graphicspath{{./figures/}}
  \DeclareGraphicsExtensions{.pdf,.jpeg,.png}

  \usepackage[figure,table]{hypcap}

\else 
  \usepackage[dvips,
    bookmarksnumbered, 
    pdfstartview=XYZ 
  ]{hyperref}  

  \usepackage[dvips]{epsfig,graphicx}
  \graphicspath{{./images/}}
  \DeclareGraphicsExtensions{.eps}

  \usepackage[figure,table]{hypcap}
\fi

\hypersetup{
    colorlinks,%
    citecolor=black,%
    filecolor=black,%
    linkcolor=black,%
    urlcolor=black
}

\title{\papertitle}

\oneauthor
  {\firstauthor}{University of Oslo \\ %
    {\tt \href{mailto:joachipo@uio.no}{joachipo@uio.no}}}

\begin{document}
\capstartfalse
\maketitle
\capstarttrue
\begin{abstract}
Birds produce multiple types of vocalizations that, together, constitute a vocal repertoire. For some species, the repertoire size is of importance because it informs us about their brain capacity, territory size or social behaviour. Estimating the repertoire size is challenging because it requires large amounts of data which can be difficult to obtain and analyse. From birds vocalizations recordings, songs are extracted and segmented as sequences of syllables before being clustered. Segmenting songs in such a way can be done either by simple enumeration, where one counts unique vocalization types until there are no new types detected, or by specific algorithms permitting reproducible studies. In this paper, we present a specific automatic method to compute a syllable distance measure that allows an unsupervised classification of bird song syllables. The results obtained from the segmenting of the bird songs are evaluated using the Silhouette metric score.
\end{abstract}

\section{Introduction}\label{sec:introduction}
According to \citeauthor{krebs1980repertoires} \cite{krebs1980repertoires}, bird vocalizations can be divided into five general categories: elements, syllables, phrases, calls and songs. These elements can be regarded as elementary sonic units in bird vocalizations. The syllables include one or more elements and are usually to a few hundred milliseconds in duration. The phrases are short groupings of syllables, while the calls are generally compact sequences of phrases. Songs, on the other hand, are long and complex vocalizations.

\begin{figure}[t]
    \centering
    \includegraphics[width=0.6\columnwidth]{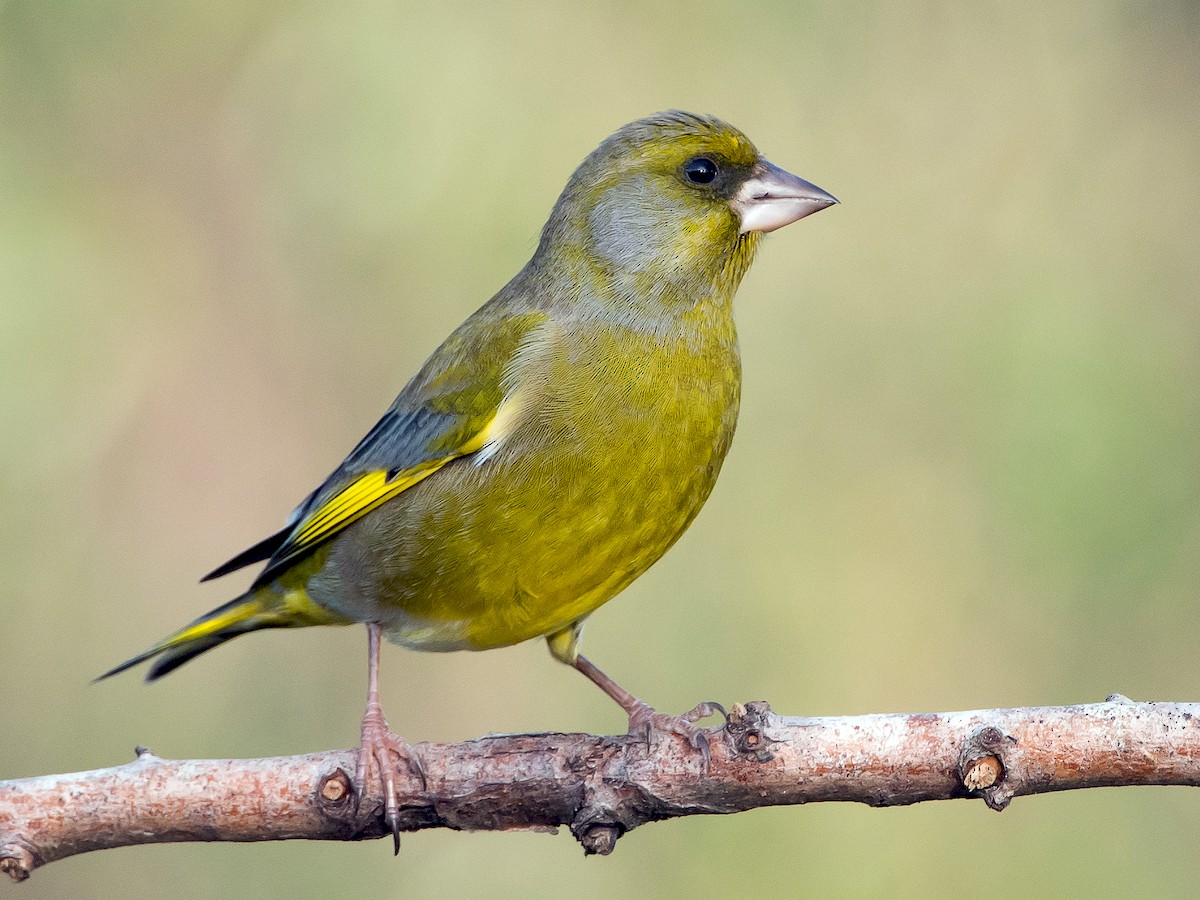}
    \caption{European Greenfinch © Rogério Rodrigues}
    \label{fig:greenfinch}
\end{figure}

The song of the European Greenfinch (Chloris Chloris) is organized in a variation of elements which may go on for more than one minute. More precisely, the song of a male greenfinch has been characterized by having groups of tremolos, repetitions of tonal units, nasal chewlee and a buzzing nasal wheeze, which could be uttered on its own \cite{cramp1994vol} and categorized into four phrases classes \cite{10.2307/4533805}: 

\begin{enumerate}
  \item A trill
  \item A pure tone
  \item A nasal tone
  \item A nasal “tswee”
\end{enumerate}

It is the “tswee” sound (a rough element, more specifically a vibrato) that characterizes the song of the greenfinch and which is an element repeated 10\% of the time \cite{shiovitz032}\cite{10.2307/4533896}. Some researchers consider the nasal “tswee” not to be learnt, but already being present in the bird’s genetic makeup. Most of the analysis found in \cite{10.2307/4533896} focuses on other elements and how they are combined (ex. the silent intervals between syllables in tours, length of tours, intervals between tours, size of the repertoire, the distance of syllables in tours among other acoustic characteristics).\\

Moreover, \citeauthor{10.2307/4533805} \cite{10.2307/4533805} argues that the size of the repertoire can be determined by the number of phrases, where, most phrases of the greenfinch, or short group of syllables, are repeated after identical intervals, for a period, on the average, which lasts 0.5 seconds. Based on these observations, we propose to use Machine Learning techniques to design an unsupervised system able to estimate the size of the repertoire of the greenfinch. The proposed system receives as input a set of audio time series data downloaded from an online collaborative database\footnote{https://xeno-canto.org} which is segmented and converted into a reduced representation set called a feature vector. Feature vector has the ability of discriminating among classes and is used to characterize the size of the bird song repertoire. The system is finally evaluated using clustering performance metrics to find the ideal number of syllables in the data set.

\section{Related Work}
\label{sec:related_work}

Estimating the size of the repertoire can be quite challenging as it needs to perform syllable classification from audio recordings. While experts can manually annotate bird syllables using simple enumeration techniques (i.e. counting the number of types present in a sample of signal) \cite{Acevedo2009AutomatedCO}, this becomes more challenging when the repertoire size is large, because counting all syllables requires large samples and a large investment of time and effort \cite{10.2307/2460394}. Previous studies used simple enumeration techniques, as well as curve-fitting \cite{wildenthal1965structure}, and capture–recapture analysis \cite{catchpole2003bird} to estimate the repertoire size. One of the bottle necks of these techniques is the segmentation of bird vocalisations into individual syllables. Simple segmentation in time domain proves difficult because of overlapping signals over different frequency bands. The common approach is to convert audio recordings into a spectrogram and apply image processing techniques to pick out the signal of interest \cite{koops2015automatic}.\\

Segmented audio signals of a specific bird species can be parameterized by a feature vector that can help discriminate between the syllables of a specific bird. Although single feature vector used for parametrization of bird song such as Mel-Frequency Cepstral Coefficients (MFCCs) \cite{harte2013identifying}, Linear Predictive Coefficients (LPC) \cite{chu2011noise}, or wavelets \cite{tanttujuha2007wavelets} can provide good results for a small number of species \cite{somervuo2006parametric}. It is noted that with the increase in the number of species, a single feature vector is not enough to be able to deal with the large diversity of sounds that different species can produce. In the same way, we assume that a similar observation can be made with the increase in the number of syllables in bird vocalization. Therefore, researchers have proposed different feature vectors in order to represent all the descriptive features \cite{zhang2018automatic}\cite{bang2017evaluation}. 
The main drawback of this strategy is that it requires the computation of all candidate features during the classification stage to build the new feature space, which can be time-consuming. Therefore, selection of features is useful to select the most relevant original features as it just requires the computation of a reduced number of selected features during the classification stage \cite{kalakech2018unsupervised}. Among the feature selection techniques, we were particularly interested in those based on individual ranking. These algorithms rank the candidate features with respect to a score which measures their relevance. In the unsupervised context, selection of features is usually done using Variance and Laplacian scores \cite{he2005laplacian}.\\

\begin{figure}[h]
    \centering
    \includegraphics[width=1\columnwidth]{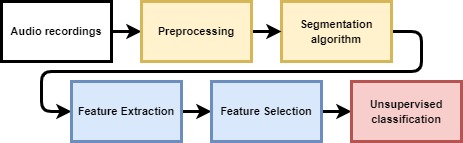}
    \caption{A schematic overview of the proposed method}
    \label{fig:pipeline}
\end{figure}

Furthermore, new estimation techniques based on automatic pattern recognition methods have been created to automate the detection of relevant structure in audio data \cite{ulloa2018estimating}\cite{ranjard2008unsupervised}. These techniques are based on unsupervised learning methods, which do not require the data to be labeled. 
Thus, the literature reveals that researchers put great efforts into producing feature vectors to discriminate birdsong as well as new estimation techniques for pattern recognition in audio data. In this study, we propose an unsupervised method (Figure~\ref{fig:pipeline}) to estimate the size of the repertoire with (1) a segmentation algorithm that extracts segments of bird audio from the recording (2) the extraction of combined vectors of descriptive features and Mel-Frequency Cepstral Coefficients, (3) the selection of features ranked with respect to a score which measures their relevance, (4) a clustering algorithm able to interpret the input data and find natural groups or clusters in the feature space. 

\section{Materials and methods}\label{sec:materials_methods}

The estimation of the size of the repertoire starts with recording the vocalization of birds. This can rarely be done in isolation, especially when the recordings are made in a natural habitat. The recordings contain not only the sound of the intended target individual but also any combination of other individuals, such as noise from other animals including humans, environmental noise (e.g. wind, water, trees, man-made noise) and electronic noise in the recorder. Parts of the audio that contains bird songs need to be segmented from the background, as well as individual syllables need to be segmented from each other before they can be used as input for the clustering algorithm.  

\subsection{Pre-processing}

In signal processing, wavelets have been widely investigated for use in filtering bio-electric signals, among many other applications. Bio-electric signals are good candidates for Multi-Resolution Analysis (MRA) \cite{mallat1989theory} wavelet filtering over standard Fourier analysis, due to their non-stationary character. Filtering of signals using wavelets is based on the idea that as the Discrete Wavelet Transform (DWT) decomposes the signal into details and approximation parts, at some scale the details contain mostly the insignificant noise and can be removed or zeroed out using threshold without affecting the signal. This idea is discussed in more detail in the introductory paper of \cite{graps1995introduction}. In this study, we propose to use a high pass filter with two basic filter design parameters: 

\begin{enumerate}
\item Wavelet type: Daubechies wavelet \cite{zhu2002image}
\item Threshold: High part of the decomposition filter values in order to remove the background noise.
\end{enumerate}

\subsection{Segmentation}

\begin{figure}[t]
    \centering
    \includegraphics[width=1\columnwidth]{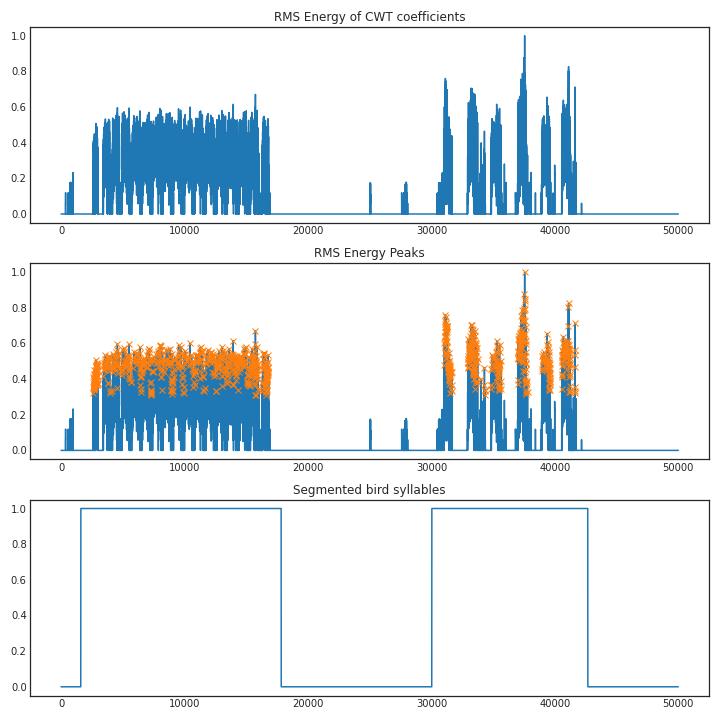}
    \caption{Segmentation of bird syllables with high-energy peaks detection}
    \label{fig:segments}
\end{figure}

Segmentation typically refers to the process of partitioning a given
document into multiple segments with the goal of simplifying the representation into something that is more meaningful and easier to analyze than the original document \cite[Chapter~4.1.1]{muller2015fundamentals}. In this study, Continuous Wavelet Transform (CWT) coefficients are calculated as inputs to the automatic segmentation algorithm as, unlike to the Fourier transform, the wavelet transform does not require to use a window function to avoid discontinuities, since wavelets are not continuous functions. Fourier Transform is used to extract frequencies from a signal as it uses a series of sine waves with different frequencies to analyze a signal. However, the main difficulty is to find the right window size. According to Heisenberg’s uncertainty principle: a narrow window will localize the signal in time but there will be significant uncertainty in frequency. If the window is wide enough, the time uncertainty increases. This is referred to as the tradeoff between time and frequency resolution. As mentioned above, one way to avoid this problem is to use MRA in order to analyze the signal at different resolution levels. In this study, we use the free library for the Python programming language \(PyWavelets\)\footnote{https://pypi.org/project/PyWavelets} to compute the CWT coefficients and calculate the energy envelope of each wavelet vector using Root Mean Square (RMS) energy function. That way, we isolate segments by finding high-energy peaks in the energy envelope and apply a threshold mask set to -20 dB (Figure~\ref{fig:segments}).

\subsection{Feature Extraction}

Feature extraction is the process in which input data are converted into a reduced representation set called a feature vector. Feature vector has the ability of discrimination among classes. In this study, the features are computed on short-term frames \(F = f1, f2, f3 . . . fn\) and these
frames are based on overlapped samples \(S = s1, s2, s3 . . . sn\) of audio signals \cite[Chapter~6.1.2]{schuller2013intelligent}. We transform the discrete signals into \(N\) short-term frames of overlapped samples and then extract the audio features from these frames.

\subsubsection{Descriptive features (DFs)}

Spectral characteristics of different birds are varied and the signal model is not known. Since bird sounds are musical in nature, time and frequency-based features, called Descriptive Features, used in audio and music retrieval can be used for bird species recognition \cite{bang2017evaluation}. These features are extensively used and described by \cite[Chapter~3]{knees2016music} and \cite[Chapter~6.2.2]{schuller2013intelligent} for music and audio retrieval, and \cite{somervuo2006parametric} and \cite{fagerlund2007bird} for recognition of bird species. In this study, the following features are used.

\begin{enumerate}
    \item Energy (EN) 
    \item Zero Crossing Rate (ZCR)  
    \item Duration of the Syllable (DUR)
    \item Spectral Centroid (SC) 
    \item Spectral Bandwidth (SB)  
    \item Spectral Flux (SF)  
    \item Spectral Roll Off (SR)  
    \item Spectral Flatness (SF)  
\end{enumerate}
 
Except the duration (DUR), all the features are extracted on frame basis, and mean (\(m\)) and variance (\(v\)) of these features are computed over the entire syllable. This gives 14 features to which duration of the syllable is concatenated, thus yielding a feature vector of length 15.

\subsubsection{Mel-Frequency Cepstral Coefficients (MFCCs)}

Mel frequency cepstral coefficients (MFCCs) have their origin in speech processing but were also found to be suited to model timbre in music. The MFCC feature is calculated in the frequency domain, derived from the signal’s spectrogram \cite[Chapter~3.2.1]{knees2016music}. In this study, we use the MFCC FB-24 configuration proposed by Cambridge Hidden Markov Models (HMM) Toolkit known as HTK because of its wide use. The name HTK MFCC FB-24 denotes the number of filters recommended by Young for speech bandwidth [0, 8000] Hz. Owing to its widespread use, MFCC filter parameters are considered as the basis for the evaluation of other feature sets. For comparison, in this study we use 24 band filters, 13 cepstral coefficients for all feature extraction methods as well as a frequency range of [0 11025] Hz as we are dealing with bird songs which are higher in frequency than speech.

\subsection{Classification}
 
\begin{figure}[h]
    \centering
    \includegraphics[width=1\columnwidth]{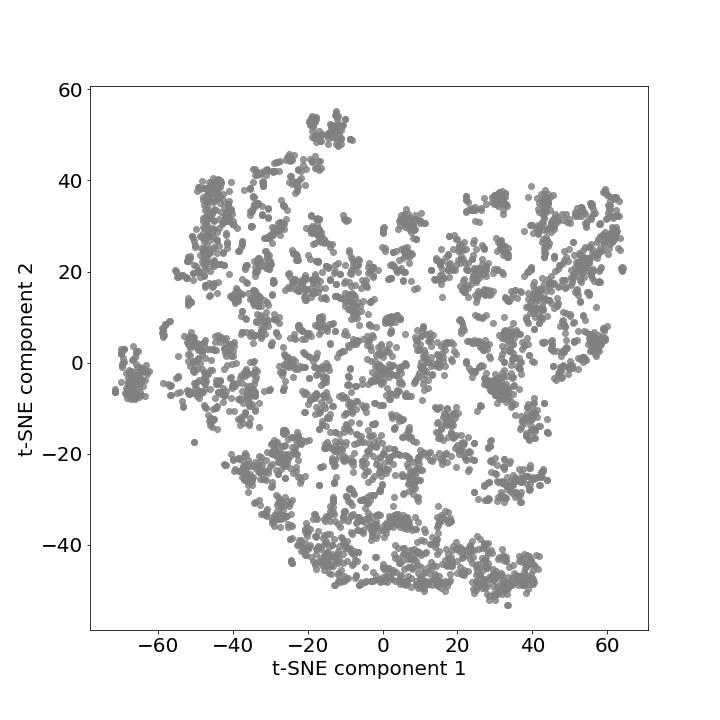}
    \caption{Exploratory visualization using the t-SNE algorithm}
    \label{fig:tsne}
\end{figure}

\begin{figure}[h]
    \centering
    \includegraphics[width=1\columnwidth]{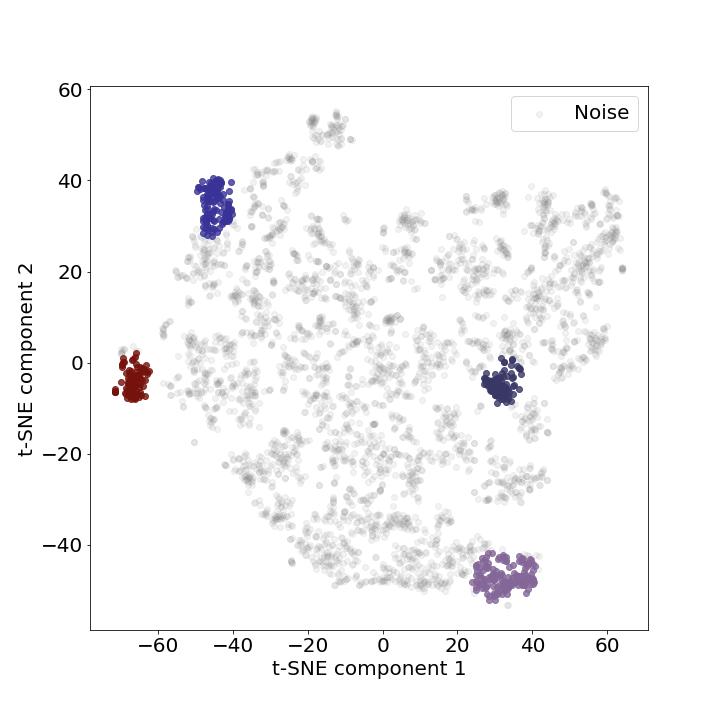}
    \caption{Unsupervised bird song syllable classification using the DBSCAN algorithm}
    \label{fig:dbscan}
\end{figure}

After features are extracted from augmented data, they are normalized using the max–min method, selected based on individual ranking, and fed into an unsupervised algorithm to automatically cluster bird syllables in the audio recordings. Because we are dealing with high-dimensional feature vector, we facilitate the classification process by applying a non-metric dimensionality reduction technique, namely the t-Distributed Stochastic Neighbor Embedding (t-SNE)\cite{van2008visualizing}, to project the data in two dimensions (Figure~\ref{fig:tsne}). Additionally, we group the samples rapidly and objectively using the DBSCAN algorithm \cite{ester1996density}. This algorithm is useful to find core samples with high density and expand clusters from them. Moreover, one of the significant attributes of this algorithm is noise cancellation which is helpful to discard th noisy samples as well as the capacity to find the number of clusters while coping with unbalanced classes (Figure~\ref{fig:dbscan}).

\section{Dataset}\label{sec:dataset}

The data set used to develop and validate the system is created using the Xeno-Canto database\footnote{https://xeno-canto.org/}, a collaborative project dedicated to sharing bird sounds from all over the world. We use Xeno Canto's web Application Programming Interface (API v2)\footnote{https://xeno-canto.org/explore/api} to build the data set with data being organized according to the API documentation. Data can be used without restrictions with a rate limit of 10 requests per second and are accessible by sending query parameters (i.e. query and page) which return a JSON object containing details about the recordings found with the given query. 
In this study, we use an area-based query gathering European recordings of the greenfinch. We select only high quality recordings according to the Xeno-Canto quality ratings ranging from A (highest quality) to E (lowest quality)\footnote{https://xeno-canto.org/api/2/recordings?query=chloris+area:europe+q:a} and remove recordings that have an other species referenced in the background. This allows us to build a data set with 339 audio recordings of an average duration of 48.64 seconds each. Audio recordings are accompanied with a detailed description of the fields of the object present in the recordings array. The following fields are kept for the study.

\begin{itemize}
    \item \textbf{id:} catalog number of the recording on xeno-canto
    \item \textbf{gen:} generic name of the species
    \item \textbf{en:} English name of the species
    \item \textbf{cnt:} country where the recording was made
    \item \textbf{file-name:} original file name of the audio file
    \item \textbf{type:} the sound type of the recording (e.g. 'call', 'song', etc)
    \item \textbf{length:} the length of the recording in minutes
\end{itemize}

\section{Feature selection}\label{sec:feature_selection}

Feature selection is a way of selecting the subset of the most relevant features from the original features set by removing the redundant, irrelevant, or noisy features. In this study, we compare two feature selection method: (1) Variance Threshold, and (2) Laplacian score, and evaluate them with the DBSCAN clustering algorithm. We select potential set of features by removing candidate features below the median score as it benefits from reduction in memory and computations, and can improve generalization and interpretability \cite{lindenbaum2021differentiable}.\\

Selection of feature vectors are generated in four phases:
\begin{itemize}
    \item Calculation of the Descriptive Features and MFCCs
    \item Normalization of the features values adjusted to the same dynamic range \((0,1)\) so that each feature has equal significance to the classification result
    \item Ranking of candidate features with respect to a score which measures their relevance
    \item Removing of candidate features below the median score to fasten the calculation
\end{itemize}

\subsection{Variance Threshold}

The Variance Threshold is a simple baseline approach to feature selection. It removes all features whose variance does not meet some threshold. By default, it removes all zero-variance features, i.e. features that have the same value in all samples. For each feature, \(f\), the Variance is computed as follows

\begin{equation}
    Var[f] = p(1 - p)
    \label{eq:VAR}
\end{equation}

\begin{figure}[t]
    \centering
    \includegraphics[width=1\columnwidth]{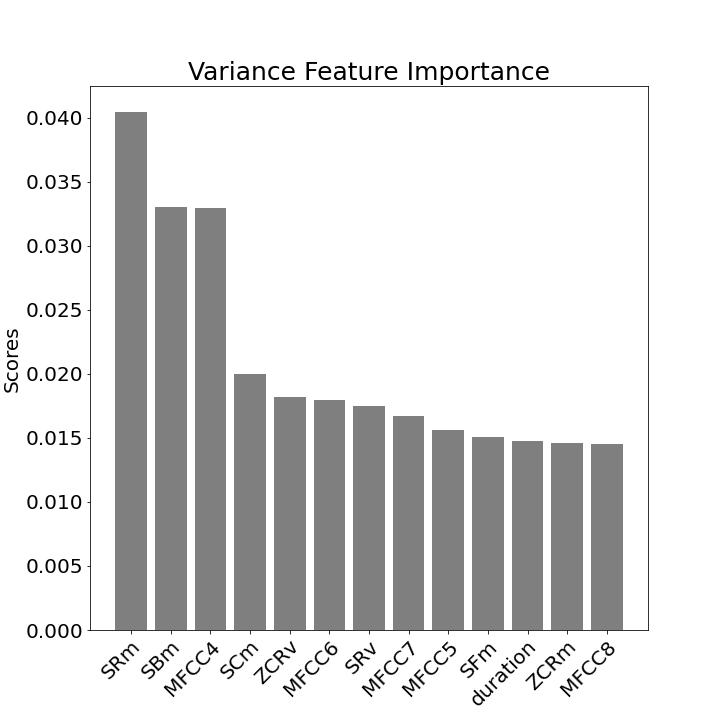}
    \caption{Variance Feature Importance scores in descending order}
    \label{fig:variance}
\end{figure}

The features with the highest \(Var[f]\) values are ranked by order of importance in Figure~\ref{fig:variance}.

\subsection{Laplacian Score}

The Laplacian score of a feature indicates its relevance to preserve locality. The Laplacian score is based on the observation that two instances that are close to each other generally belong to the same class. Laplacian score uses the nearest neighbor graph to obtain the local structure of the data and obtains the Laplacian score value of each feature according to the distance metric (euclidean).

\begin{figure}[t]
    \centering
    \includegraphics[width=1\columnwidth]{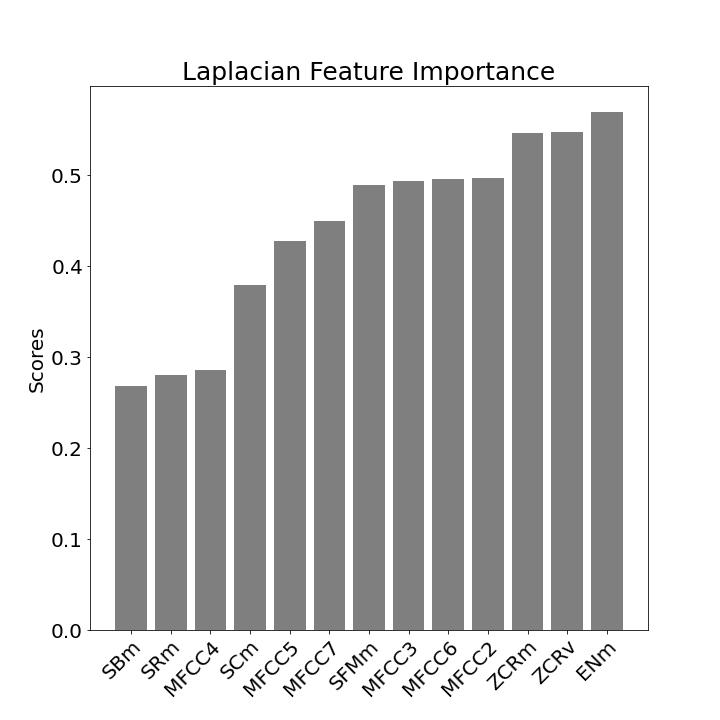}
    \caption{Laplacian Feature Importance scores in ascending order}
    \label{fig:laplacian}
\end{figure}

For each feature, \(f\), the Laplacian score of the \(r\)th feature is computed as follows

\begin{equation}
    L_r = \frac{\tilde{f}_{r}^T L\tilde{f}_r}{\tilde{f}_{r}^T D\tilde{f}_r}
    \label{eq:LPS}
\end{equation}

where the vector \(\tilde{f}_{r}\) is

\begin{equation}
    \tilde{f}_{r} = {f}_{r} - \frac{{f}_{r}^T D1}{1^T D1}
    \label{eq:LPS1}
\end{equation}

and \(L\) and \(D\) are defined in the spectral algorithm. Based on the assumption that the interesting
underlying structure of the data (e.g. clusters) depends on the slowly varying features in the data, \cite{he2005laplacian} proposed to select the features with the smallest scores. The features are ranked by order of importance in Figure~\ref{fig:laplacian}.\\

\section{Results and discussion}\label{sec:results}

After the selection of the feature vectors, we use the free software machine learning library for the Python programming language \textit{scikit-learn} \cite{pedregosa2011scikit} to implement the unsupervised clustering techniques. We start by projecting the feature vectors in two dimensions using the t-SNE algorithm and save the following parameters as they represent the best visualization of the data.

\begin{itemize}
    \item \textbf{perplexity} = 45
    \item \textbf{n\_iter} = 1500
    \item \textbf{init} = 'pca'
    \item \textbf{learning\_rate} = 'auto'
\end{itemize}

We then adjust the main parameters of the DBSCAN clustering algorithm (i.e. \(eps\) and \(min\_samples\)) using k-distance graph to find the optimal epsilon (\(eps\)) distance between the samples, and Silhouette metric \cite{rousseeuw1987silhouettes} to determine the minimum number of samples (\(min\_samples\)) needed to define the optimal number of clusters. 
Observations showed that there was never more than one big cluster. At most, there was one large cluster identified as noise and some small clusters identified as bird song syllables. 

\subsection{Evaluation}

We evaluate the proposed system using the DBSCAN algorithm and the Silhouette metric score with euclidean distance calculation. Silhouette score always ranges between -1 to 1 with a high score suggesting that the objects are well matched to their own cluster and poorly matched to their neighborhood clusters. This is computed as follows. 

\begin{equation}
    s = \frac{b - a}{max(a, b)}
    \label{eq:silhouette}
\end{equation}
Where:
\begin{description}
\item\(a\) = mean distance between a sample and all other points in the same class
\item\(b\) = mean distance between a sample and all other points in the next nearest cluster
\end{description}

\subsubsection{Euclidean distance method}

The biggest challenge with the DBSCAN algorithm is to find the right parameters to model the algorithm. According to \cite{ester1996density}, the minimum number of samples in a neighborhood for a point to be considered as a core point should be greater than or equal to the dimensionality of the dataset multiply by 2. In this study, we start setting DBSCAN’s default \(min\_samples\) value to 26 as our feature vector contains 13 dimensions.

\begin{figure}[t]
    \centering
    \includegraphics[width=1\columnwidth]{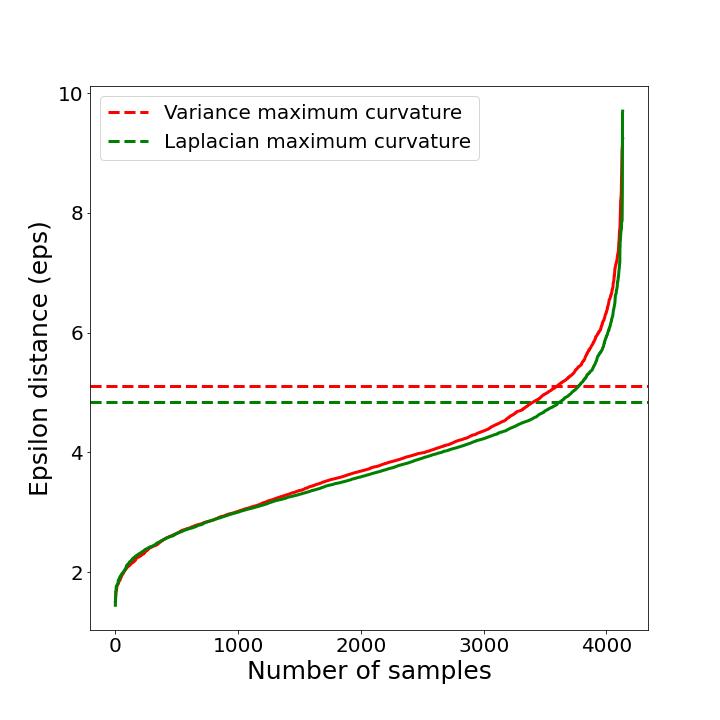}
    \caption{Optimal epsilon distance (\(eps \approx5\))}
    \label{fig:epsilon}
\end{figure}

We then calculate the minimum euclidean distance among each data points and plot the result (Figure~\ref{fig:epsilon}). The ideal value for \(eps\) is equal to the distance value at the "knee" point, or the point of maximum curvature. This point represents the optimization point where diminishing returns are no longer worth the additional cost. This concept of diminishing returns applies here because while increasing the number of clusters will always improve the fit of the model, it also increases the risk that over fitting will occur. We compute the knee point of our function using the free library for the Python programming language \(kneed\)\footnote{https://pypi.org/project/kneed}.

\begin{figure}[t]
    \centering
    \includegraphics[width=1\columnwidth]{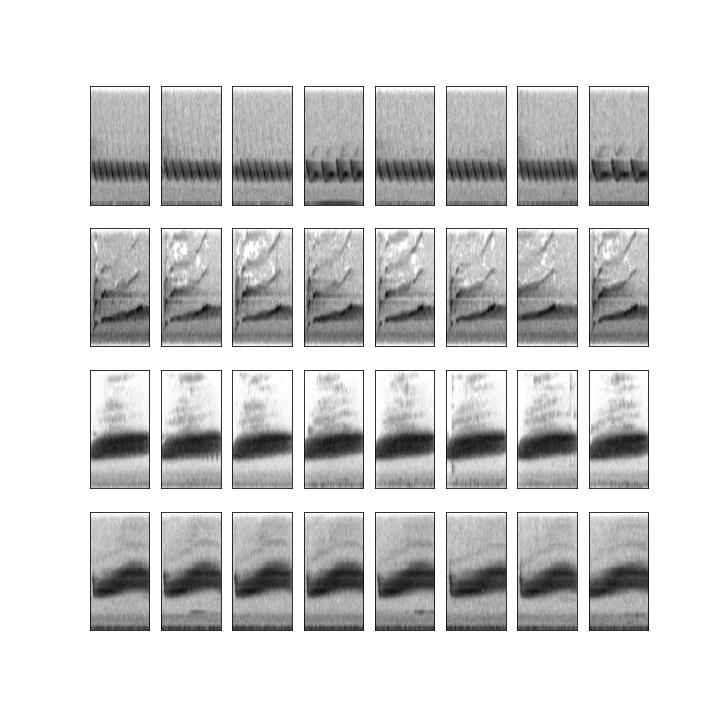}
    \caption{Spectrograms of the four clusters of syllables}
    \label{fig:syllable}
\end{figure}

Finally, we set the DBSCAN algorithm with the optimal \(eps\) parameter through various ranges of minimum samples values (i.e. from 20 to 75). We find the right value according to the highest Silhouette score while comparing the two selected feature vectors. From Table \ref{tab:results}, we can see that \(min\_samples\) value should be set to 75 as it yields the number of clusters to the highest Silhouette score. Clustering performance evaluation isolate five clusters using Laplacian feature vector selection with four clusters of bird syllables (Figure~\ref{fig:syllable}) for a maximum Silhouette score of 0.88\% and one cluster of outliers.

\begin{figure}[t]
    \centering
    \includegraphics[width=1\columnwidth]{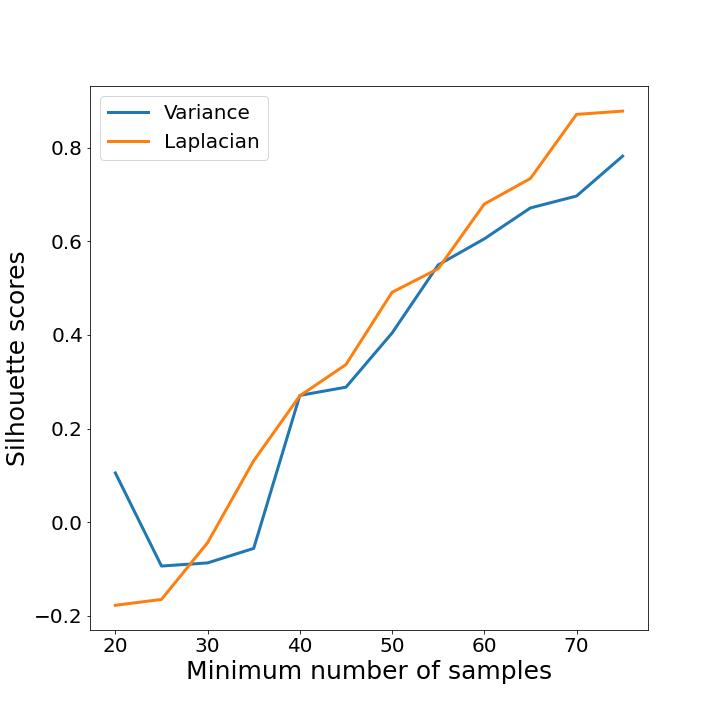}
    \caption{Clustering performance evaluation based on the Silhouette metric}
    \label{fig:silhouette}
\end{figure}

\begin{table}[t]
\centering
\begin{tabular}{|c|c|c|}
    \hline
    \multicolumn{3}{|c|}{\textbf{Variance | Laplacian}}\\
    \hline
    min\_sample & Clusters found & Silhouette scores\\
    \hline
    20 & 1 | 4 & 0.105 | -0.177\\ 
    \hline
    25 & 4 | 4 & -0.093 | -0.165\\ 
    \hline
    30 & 4 | 11 & -0.086 | -0.043\\
    \hline
    35 & 10 | 15 & -0.055 | 0.131\\ 
    \hline
    40 & 14 | 24 & 0.270 | 0.269\\
    \hline
    45 & 21 | 25 & 0.288 | 0.336\\ 
    \hline
    50 & 26 | 27 & 0.403 | 0.491\\
    \hline
    55 & 22 | 20 & 0.549 | 0.541\\ 
    \hline
    60 & 19 | 14 & 0.605 | 0.679\\
    \hline
    65 & 12 | 11 & 0.671 | 0.734\\ 
    \hline
    70 & 10 | 4 & 0.696 | 0.870\\
    \hline
    \textbf{75} & \textbf{8} | \textbf{4} & \textbf{0.781} | \textbf{0.878}\\
    \hline
\end{tabular}
\caption{Silhouette scores for minimum sample values}
\label{tab:results}
\end{table}

\section{Conclusions}
As mentioned by \citeauthor{10.2307/4533805} in \cite{10.2307/4533805}, the size of the repertoire can be determined by the number of its phrases (short groupings of syllables). In this paper, we estimated the repertoire size of the greenfinch based on bird song data from different places in Europe. The method is based on (1) a
segmentation algorithm that extracts segments of bird audio from the recording, (2) the extraction and (3) the selection of combined vectors of audio features. This study provided us with a repertoire size estimation distinguishing four clusters of syllables using unsupervised clustering techniques. 

Most phrases of the greenfinch have been characterized into four classes by \citeauthor{10.2307/4533805}: 

\begin{enumerate}
  \item A trill
  \item A pure tone
  \item A nasal tone
  \item A nasal “tswee”
\end{enumerate}

Based on the scores of the clustering performance evaluation of our system, we found a similar number of classes. This confirms \citeauthor{10.2307/4533805}’s results despite the fact that his method is different from the one employed in this study. 

The main benefits of the system are the visualization and the sonification of bird syllables clusters which turns out to be accurate for finding the presence of similar patterns in the data. Nevertheless, due to the strong presence of noisy samples detected by the DBSCAN algorithm in the data (3678 out of a total of 4133 samples), the proposed system can under or overestimate the repertoire size in birds. Consequently, this system is only useful for an experimental estimation of the repertoire size. Further improvements can focus on finding more robust techniques to denoise the audio recordings such as denoising auto-encoder neural networks as discussed in \cite{stowell2015denoising} or experimenting new feature extraction techniques such as those discussed in \cite{ulloa2018estimating}. 

Furthermore, geographic variation between the greenfinches songs from Spain, France, Germany, Denmark, Britain and New Zealand have been found in \cite{10.2307/4533805}. Thus, new study can explore whether geographic variation is also observed in the data that has been selected, by focusing on the specific syllable clusters found by the unsupervised algorithm. However, selection of the data will have to be rethought in order to homogenize the number of recordings for each European country.

\subsection*{Data Accessibility}
The audio recordings are accessible from the Xeno-Canto sound library (\url{https://xeno-canto.org}). Source codes (Python) are available at: \url{https://github.com/joachimpoutaraud/estimating-repertoire-size-in-birds}. A step by step instruction to run the study is provided in the notebooks of the repository.

\printbibliography

\end{document}